\documentclass[conference]{IEEEtran}
\usepackage[utf8]{inputenc}
\usepackage{graphicx}
\usepackage{amsmath}
\usepackage{hyperref}

\title{Can AI Have a Personality? Prompt Engineering for AI Personality Simulation: A Chatbot Case Study in Gender-Affirming Voice Therapy Training}

\author{
	\IEEEauthorblockN{Tailon Jackson\IEEEauthorrefmark{1}, Dr. Byunggu Yu\IEEEauthorrefmark{2}} \\
	\IEEEauthorblockA{\IEEEauthorrefmark{1}University of the District of Columbia\\
		Email: tailon.jackson@udc.edu} \\
	\IEEEauthorblockA{\IEEEauthorrefmark{2}University of the District of Columbia\\
		Email: byu@udc.edu}
}

\begin{document}
	
	\maketitle
	
	\begin{abstract}
		This thesis investigates whether large language models (LLMs) can be guided to simulate a consistent personality through prompt engineering. The study explores this concept within the context of a chatbot designed for Speech-Language Pathology (SLP) student training, specifically focused on gender-affirming voice therapy. The chatbot, named Monae Jackson, was created to represent a 32-year-old transgender woman and engage in conversations simulating client-therapist interactions. Findings suggest that with prompt engineering, the chatbot maintained a recognizable and consistent persona and had a distinct personality based on the Big Five Personality test. These results support the idea that prompt engineering can be used to simulate stable personality characteristics in AI chatbots.
	\end{abstract}
	
	\section{Introduction}
	
	Can AI Have a Personality? Prompt Engineering for AI Personality Simulation: A Chatbot Case Study in Gender-Affirming Voice Therapy Training 
	
	This thesis investigates whether large language models (LLMs) can be guided to simulate a consistent personality through prompt engineering. The study explores this concept within the context of a chatbot designed for Speech-Language Pathology (SLP) student training, specifically focused on gender-affirming voice therapy. The chatbot, named Monae Jackson, was created to represent a 32-year-old transgender woman and engage in conversations simulating client-therapist interactions. 
	
	The research is grounded in a single hypothesis: that carefully constructed prompts can influence an AI chatbot to behave as if it has a stable, predefined personality. Monae’s behavior was shaped through detailed persona development and iterative prompt refinement. The prompts were designed to encourage consistency in tone, emotional response, and conversational boundaries and to guide the chatbot’s conversational behavior throughout each session.  
	
	To evaluate the chatbot’s responses, the Big Five Personality test was used to compare the chatbot's simulated responses to expected personality traits and the Jungian/Myers Briggs personality assessment was used to classify Monae’s personality and contrast the known personality traits of ChatGPT. Findings suggest that with prompt engineering, the chatbot maintained a recognizable and consistent persona and had a distinct personality based on the Big Five Personality test.   
	
	These results support the idea that prompt engineering can be used to simulate stable personality characteristics in AI chatbots. While the scope of the study is limited, the findings provide a starting point for further research into the use of AI personas in educational and therapeutic simulations. Future work could explore more complex personality traits, expand the number of personas, and test effectiveness across different clinical education contexts. Broader evaluations may help determine whether such tools can supplement human-led training in a consistent and ethically responsible manner. 
	
	\subsection{Summary}
	
	This thesis investigated whether structured prompt engineering could be used to simulate a coherent and consistent AI personality across multiple interactions and assessments. The project focused on the development of Monae Jackson, a chatbot persona created to support Speech-Language Pathology (SLP) students in practicing gender-affirming voice therapy. The study addressed a specific training gap by exploring whether large language models (LLMs) could embody personality traits through the use of detailed character background prompts and behavioral constraints. 
	
	The primary objective, assessing whether prompt engineering enables the consistent simulation of personality, was supported through and quantitative evaluation. Monae’s personality was tested using two separate psychometric frameworks: the Big Five Personality Test and a Jungian-based typology modeled after the Myers-Briggs Type Indicator. Despite being administered in independent sessions, the chatbot produced highly similar results across both assessments, with strong alignment in emotional sensitivity, introversion, openness, and interpersonal warmth. Notably, consistent personality traits not explicitly programmed, such as messiness and artistic interest, emerged independently across both tests. These emergent behaviors suggest that carefully designed prompts can guide LLMs not only to adopt predefined behaviors but also to generalize in realistic and psychologically plausible ways.

	\section{Background} 
	
	Artificial Intelligence (AI) has increasingly become integrated into fields such as healthcare, education, and communication. Within these domains, AI systems offer new possibilities for delivering services, personalizing learning, and supporting complex tasks that previously relied solely on human expertise. The use of conversational AI, particularly chatbots, is one area that has drawn growing attention due to its accessibility, scalability, and ability to simulate human-like interactions. 
	
	Among recent advancements, large language models (LLMs) such as OpenAI’s GPT family represent a major leap in natural language generation. These models are capable of producing coherent, contextually aware responses, making them suitable for applications requiring extended human-like dialogue. Their utility spans a wide range of tasks, from tutoring systems to virtual assistants and therapeutic simulations. As these systems evolve, questions about their capacity to replicate more complex human behaviors—such as emotional consistency, tone, or personality, have become increasingly relevant, Yu and Kim (2023). 
	
	Efforts to simulate human conversation date back to early systems such as ELIZA (Weizenbaum, 1966), a simple rule-based program that mimicked a Rogerian psychotherapist. Despite its limitations, ELIZA demonstrated how users might attribute emotional or personal qualities to even rudimentary AI systems. Over the years, improvements in computational power and machine learning led to more advanced dialogue systems, such as Microsoft’s XiaoIce and modern large-scale transformers. 
	
	While earlier systems relied on structured scripts and decision trees, contemporary LLMs are trained on vast datasets and generate text probabilistically, allowing more flexible and naturalistic conversation. However, these models do not possess memory or intrinsic personality unless specifically prompted to simulate one. The idea that personality traits can be imposed or maintained through the input text, what is now often referred to as “prompt engineering”, has become a focal point in AI development (Reynolds \& McDonell, 2021).
	
	Research has shown that language models can emulate stylistic elements, emotional tone, and character-based behavior when carefully prompted. Applications in mental health, education, and virtual customer service have demonstrated that users are more likely to engage with chatbots when the agent displays empathy, coherence, and relatable traits (Bickmore \& Cassell, 2005).
	
	Despite this progress, much of the existing literature focuses on general conversational coherence or task performance, rather than evaluating whether an AI can consistently portray a specific personality. Some studies have explored alignment with psychological constructs, but there is limited empirical work examining how a chatbot's behavior aligns with structured personality assessments or remains stable over multiple sessions (Yu and Kim 2023). 
	
	Key questions remain about whether AI can not only simulate natural conversation but also embody a consistent persona over time and across contexts. Research is limited in demonstrating whether these personas can be evaluated using standard psychometric tools or whether they meaningfully influence educational outcomes. Furthermore, few studies have explored these questions within highly specialized fields such as Speech-Language Pathology (SLP), where interactions must combine technical knowledge with sensitive communication. 
	
	There is also a gap in understanding how educational simulations can benefit from AI-driven personalities, especially in the context of training students to engage with diverse or underrepresented populations such as transgender individuals receiving gender-affirming care. 
	
	This thesis addresses these gaps by exploring the use of prompt engineering to simulate a consistent AI personality within an educational chatbot. The project centers on Monae Jackson, a chatbot representing a transgender woman seeking gender-affirming voice therapy. By embedding a detailed personality through structured prompts and evaluating the resulting behavior, this research investigates whether large language models can behave in ways that resemble psychologically coherent personas. 
	
	Specifically, the study examines whether the chatbot’s behavior aligns with a predefined character profile and whether this personality remains stable across different interactions. In doing so, it contributes to the growing field of AI personality simulation and offers potential applications for SLP training and similar clinical education environments. 
	
	\subsection{Motivation}
	
	This research is motivated by an ongoing need to improve training resources within Speech-Language Pathology (SLP) education, particularly in the area of gender-affirming care. While awareness of transgender and gender-diverse identities has grown significantly in recent years, many SLP programs continue to offer limited exposure to client populations with gender-specific therapeutic needs. Voice therapy, especially when integrated as part of broader gender-affirming healthcare, plays a critical role in supporting transgender individuals in achieving greater self-alignment, confidence, and social comfort. Despite its clinical importance, opportunities for students to engage meaningfully with these types of clients remain constrained. 
	
	This thesis responds to a proposed collaboration with the SLP department at the University of the District of Columbia, initiated through Professor Carmen Ramos-Pizarro. This idea, to explore whether simulated therapeutic interactions, delivered through a carefully constructed AI chatbot persona, could help supplement traditional instruction for SLP students at the University of the District of Columbia, for interactions with gender-diverse individuals, was the sole idea and contribution from Professor Carmen Ramos-Pizarro. Professor Carmen Ramos-Pizarro applied for the funding from the ASHA Multicultural Grant to support the project.  
	
	As the researcher, I also bring a personal interest in contributing to systems that support equitable and competent care for that same population. The aim is not merely to create a functional tool, but to examine whether personality-consistent AI systems can serve as safe, repeatable, and pedagogically meaningful practice partners. These simulations could offer SLP students the opportunity to gain confidence, make mistakes, and receive feedback in a controlled setting without the risk of causing harm to a real person due to inexperience. Ensuring that the chatbot is culturally responsive, emotionally realistic, and guided by best practices is central to this goal. In doing so, this work hopes to contribute to a broader conversation about how artificial intelligence can responsibly support healthcare education, particularly in areas where lived experience is vital but difficult to replicate in training. 
	
	\subsection{Problem Statement}  
	
	A critical shortfall in current SLP education is the inadequate exposure of students to realistic and diverse clinical interactions. The lack of sufficient opportunities to practice gender-affirming therapeutic interactions in controlled educational environments directly impacts the quality of therapeutic care provided to gender-diverse clients. 
	
	Students can often finish their entire program without having the opportunity to work with a gender-diverse client for many reasons but consistently because of the lack of clients to practice on. This is understandable, because that population is vulnerable and agreeing to help SLP students could open a gender-diverse person up to potentially hurtful and negative interactions.  
	
	Without access to authentic scenarios and detailed feedback on interpersonal interactions, clinicians may be unprepared for real-world therapeutic settings, potentially affecting their effectiveness and ability to build trust and rapport with clients. Inadequate exposure may lead to unintended insensitivity during therapy sessions, negatively impacting the therapeutic relationship and outcomes. 
	
	Addressing this gap in educational preparation requires the exploration of scalable and accessible solutions. AI-driven chatbots present a potential method for simulating realistic and emotionally complex clinical interactions, particularly in training contexts where direct experience may be limited. By enabling repeated, low-risk practice with a consistent virtual persona, such tools may support the development of both clinical communication skills and the interpersonal sensitivity necessary for working with gender-diverse client populations. 
	
	\subsection{Research Objectives}
	
	To address these challenges comprehensively, this thesis outlines the following objectives: 
	
	\begin{itemize}
		\item Assess the viability of embedding consistent and detailed personalities within AI-driven chatbots through advanced prompt engineering techniques.
		\item Implement and evaluate a realistic chatbot persona designed specifically for gender-affirming voice therapy training scenarios.
		\item Evaluate the chatbot’s personality consistency and depth through psychometric tests, specifically employing the Big Five Personality test and the Jungian/Myers Briggs personality test. 
	\end{itemize}
	
	By pursuing these objectives, this study seeks to examine whether AI chatbots, when guided by structured prompt engineering, can simulate consistent and contextually appropriate interactions that support training in therapeutic education settings. 
	
	\section{Thesis Outline}  
	
	This thesis systematically addresses the above objectives across several chapters. Following this introductory chapter, Chapter 2 provides a comprehensive literature review exploring foundational concepts of AI personality, gender-affirming therapeutic methodologies, prompt engineering techniques, and ethical considerations. Chapter 3 details the methodological approach, describing the technical infrastructure, persona development, prompt engineering, and evaluation strategies employed. 
	
	Chapter 4 covers the implementation process, discussing application development, hosting considerations, user interface design, and logging mechanisms. Chapter 5 presents the research results, including performance metrics, consistency, and personality assessment outcomes. Chapter 6 offers an in-depth discussion of findings, educational implications, challenges encountered, and potential limitations of the chatbot system. Future directions are outlined in Chapter 7 which also concludes the thesis, summarizing key insights and the potential impact of AI chatbots in educational and therapeutic training settings.

	\section{Literature Review}
	
	\subsection{Foundations of AI Personality}
	
	The concept of imbuing artificial intelligence with personality has deep historical roots, originating from early discussions around Alan Turing's proposed test for artificial intelligence. The Turing Test originally questioned whether machines could exhibit behavior indistinguishable from human interactions (Weizenbaum, 1966). Over the decades, as computing capabilities and algorithmic complexity grew exponentially, this foundational idea evolved, enabling AI systems to demonstrate increasingly sophisticated human-like interactions. 
	
	Early conversational AI models such as ELIZA, created by Joseph Weizenbaum in 1966, laid important groundwork. ELIZA was designed to simulate a Rogerian psychotherapist through simple text-based interactions, thereby highlighting the human propensity to attribute personality traits and emotional understanding to conversational agents, even those with very rudimentary programming (Weizenbaum, 1966). 
	
	Advancements in Large Language Models (LLMs) such as GPT-4 have significantly elevated the capabilities of conversational AI. These modern models can handle complex language constructs, context awareness, and dynamic conversational threads, far surpassing early systems (Brown et al., 2020). Despite these advancements, challenges remain in ensuring consistent and ethically appropriate AI personality presentations. The models are trained on vast datasets from diverse and sometimes inconsistent sources, potentially embedding biases or stereotypical behavior (Weidinger et al., 2021). Therefore, careful and deliberate prompt engineering is critical to define clear behavioral guidelines for AI personas (Reynolds \& McDonell, 2021).

	Research by Bickmore and Cassell (2005) emphasizes that carefully constructed conversational interfaces can significantly enhance user engagement and trust by simulating social dialogue and emotional responsiveness. Recent studies continue to investigate the subtle yet powerful effects that clearly defined AI personalities can have on user experience, therapeutic efficacy, and educational outcomes (Yu \& Kim, 2023). 
	
	\subsection{OpenAI’s Work with ChatGPT and Personality Alignment}
	
	OpenAI’s development of ChatGPT builds on foundational work in LLM training, fine-tuning, and alignment. ChatGPT uses reinforcement learning from human feedback (RLHF) and supervised fine-tuning to improve response quality, reduce toxicity, and encourage more helpful, honest, and harmless interactions (Ouyang et al., 2022; Askell et al., 2021). These alignment processes have become known as “basic alignment” and are central to the behavior shaping of public-facing AI. 
	
	While OpenAI’s alignment research has largely focused on factuality, safety, and user preferences, emerging research highlights the role of prompt engineering in influencing perceived personality traits. As shown in the study by Yu and Kim (2023), ChatGPT can exhibit measurable personality dimensions when assessed using psychometric tools such as the Hogan Personality Inventory (HPI) and the Big Five. 
	
	Their results indicated that ChatGPT, when prompted without role instructions, exhibited low sociability, but this trait could be steered through structured role-playing prompts. These findings suggest that large language models may implicitly reflect personality-like behaviors based on training data and alignment methods, but they also indicate that these behaviors are potentially malleable. Yu and Kim (2023) evaluated ChatGPT’s responses to two standardized personality assessments, the Hogan Personality Inventory (HPI) and the IPIP-NEO-120, without applying any role-based or behavioral prompts. Their goal was to determine whether ChatGPT would express stable personality traits when answering in its default state. 
	
	The model’s responses were assessed using established scoring rubrics, and the results revealed a consistent personality profile across repeated test administrations. Specifically, ChatGPT scored lower on traits such as sociability and ambition, and higher on prudence and inquisitiveness. These findings suggest that large language models exhibit stable but implicit personality patterns, even when not explicitly prompted to perform a role and the results support further exploration into whether AI personalities can be intentionally shaped through prompt engineering or alignment techniques. 
	
	This thesis draws upon this framework to explore how personality traits can be simulated in ChatGPT through prompt engineering, using Monae Jackson as a case study. Rather than focusing on factual alignment alone, this work investigates whether consistent behavioral and emotional patterns can be maintained across interactions through structured prompts, contributing to a growing body of research on AI personality simulation and alignment.

	\subsection{Personality Theory and Assessment}
	
	The study of personality aims to understand consistent patterns of behavior, thought, and emotion that distinguish individuals. In psychology, personality is typically defined as the enduring traits and characteristics that influence how people interact with the world around them. Over time, researchers have developed several theoretical frameworks and standardized tools to measure personality in a reliable and structured way. 
	
	Two of the most widely known personality assessment frameworks are the Five-Factor Model (also known as the Big Five) and the Jungian-based typology that underlies the Myers-Briggs Type Indicator (MBTI). These models have been used in both research and applied settings, including clinical assessment, workplace evaluation, and for this study, in AI research to evaluate simulated personas.

	\subsubsection{The Big Five Personality Model}
	
	The Big Five model organizes personality into five broad traits: Openness to Experience, Conscientiousness, Extraversion, Agreeableness, and Neuroticism (often abbreviated as OCEAN). This framework has emerged from decades of lexical and statistical research, with its origins tracing back to work by researchers such as Tupes and Christal (1961) and later formalized by Costa and McCrae (1985). The Big Five model is considered one of the most empirically robust frameworks for measuring personality, supported by extensive validation across cultures, age groups, and contexts. The model’s widespread use is largely due to its reliability, predictive validity, and transparency in scoring. Assessments such as the IPIP-NEO (International Personality Item Pool) are based on this model and are publicly available, making them ideal for academic research. 
	
	\subsubsection{Jungian Theory and the Myers-Briggs Type Indicator}
	
	In contrast to the trait-based Big Five, the Myers-Briggs Type Indicator (MBTI) is based on Carl Jung’s theory of psychological types, which proposes that people differ in how they perceive the world and make decisions (Myers et al., 1998). Developed by Katharine Briggs and Isabel Briggs Myers in the 1940s, the MBTI organizes personality into four dichotomies: Introversion–Extraversion, Sensing–Intuition, Thinking–Feeling, and Judging–Perceiving. These dimensions combine into 16 possible personality “types.”
	The MBTI is widely recognized and used in business, education, and personal development, though it is often critiqued by psychologists for lacking the predictive validity and test–retest reliability seen in trait-based models like the Big Five. Despite this, MBTI remains popular due to its intuitive format, descriptive profiles, and appeal in self-reflection and team-building contexts. Variants such as the Open Extended Jungian Type Scales (OEJTS) offer a free and openly accessible alternative for researchers interested in Jungian-type assessments.

	\subsubsection{Personality Measures}
	
	For a personality assessment to be considered trustworthy, it must meet criteria such as construct validity (accurately measuring what it claims to measure), reliability (yielding consistent results over time), and predictive utility (being able to predict relevant real-world outcomes). The Big Five, particularly through validated instruments like the IPIP-NEO, consistently meets these standards and is widely accepted in academic psychology. While the MBTI and Jungian models do not offer the same empirical rigor, they remain valuable in exploratory and applied contexts where rigid scientific validation is not the primary goal.
	
	In the context of AI personality evaluation, trait-based instruments like the Big Five are particularly useful. Their dimensional scoring and open-access versions make them practical tools for evaluating whether AI-generated text exhibits stable personality patterns across time or in response to specific prompts.
	
	\subsection{Gender-Affirming Therapeutic Approaches}
	
	Gender-affirming voice therapy is recognized as an important aspect of healthcare for transgender and gender-diverse individuals. This therapeutic intervention focuses on modifying vocal attributes, such as pitch, intonation, and resonance, to align more closely with a client’s gender identity. The aim is to support improved social communication, psychological well-being, and overall quality of life (Hancock, Colton, \& Douglas, 2014; Davies et al., 2015). Treatment protocols often integrate both physiological voice training and psychosocial components, acknowledging that voice is a key factor in gender expression and social perception.
	
	The World Professional Association for Transgender Health (WPATH, 2012) has outlined guidelines emphasizing that voice therapy should be delivered within a broader framework of gender-affirming care. These guidelines recommend patient-centered, trauma-informed, and culturally competent approaches. Beyond the technical aspects of vocal training, effective therapy requires clinicians to demonstrate consistent affirming behavior, emotional sensitivity, and awareness of the sociocultural contexts that shape clients’ experiences.
	
	Given the interpersonal and emotionally complex nature of this work, clinicians must be prepared to engage with clients in ways that validate identity and reduce psychological distress. Training environments that offer opportunities to practice these interpersonal competencies, particularly through simulated interactions that reflect the realities of gender-affirming care, may help better prepare future speech-language pathologists for these responsibilities (McNeill, Wilson, \& Clark, 2008).
	
	\subsection{Techniques in Prompt Engineering}
	
	Prompt engineering is an emerging area of practice concerned with designing effective input instructions to guide the behavior of large language models (LLMs). In systems such as GPT-4o, outputs are highly sensitive to the structure and content of the prompts they receive. As such, the way prompts are phrased can significantly influence the relevance, tone, and coherence of the model’s responses (Zhou et al., 2023). Effective prompt engineering is especially important when attempting to simulate consistent persona-based interactions, where the goal is not only to retrieve information but to maintain a specific behavioral or emotional style across a conversation.
	
	Earlier approaches to prompt construction often relied on minimal or “few-shot” examples, offering brief context or representative input-output pairs. However, research by Reynolds and McDonell (2021) demonstrates that more structured and detailed prompts—sometimes called instruction-style prompting—are more effective in producing stable and role-consistent outputs. These prompts can embed persona definitions, dialogue constraints, and response style guidelines, allowing developers to exert more precise control over a model’s behavior.
	
	In this study, two advanced strategies are employed to reinforce persona consistency: recursive iteration and character background infusion. Recursive iteration refers to a continuous cycle of prompt evaluation and revision, wherein feedback from prior interactions informs the next version of the prompt. Character background infusion involves embedding biographical and emotional context about the persona into the prompt, enabling the model to generate responses that align with predefined personality traits and behaviors.
	
	These techniques are particularly valuable in settings that require emotional nuance and interpersonal sensitivity, such as educational simulations for gender-affirming voice therapy. Without such strategies, language models may exhibit response drift, lose coherence, or produce responses inconsistent with the intended personality. By contrast, prompt engineering enables the creation of more realistic, bounded, and context-aware conversational agents, an important step toward ethically deploying AI in human-centered training environments (Zhou et al., 2023; Tamkin et al., 2021).
	
	\subsection{Ethical and Practical Considerations}
	
	The integration of AI-driven chatbots in therapeutic and educational settings presents numerous ethical and practical challenges. Key concerns include data privacy, the accuracy of representation, and potential biases inherent in training datasets, all of which demand meticulous oversight. As the use of chatbots to emulate sophisticated human interactions increases, it is crucial to ensure that these simulations adhere to ethical standards and exhibit cultural sensitivity.
	
	Although AI chatbots can serve as valuable educational tools, it is important to acknowledge that they cannot fully replicate the lived experiences of real individuals, particularly those from marginalized communities. Their role in training must be clearly communicated to users to prevent overreliance or misinterpretation. Transparent documentation about the chatbot's limitations, scope, and intended use is essential to maintaining appropriate expectations.
	Additionally, the successful deployment of chatbots in education depends on practical factors such as usability, accessibility, and interaction consistency. Tools that are difficult to use, unreliable, or produce inconsistent behavior may diminish educational outcomes rather than enhance them. Systems should be evaluated not only for their technical performance but also for their capacity to model empathy, uphold ethical boundaries, and respond in culturally sensitive ways.
	
	Addressing these ethical and practical considerations requires a comprehensive, multidisciplinary approach, involving continuous collaboration among AI developers, educators, clinical experts, and community representatives. This collaboration helps to ensure AI simulations are respectful, accurate, and beneficial for sensitive educational and therapeutic scenarios.

	\section{Methodology}
	
	\subsection{Technical Infrastructure}
	
	The technical infrastructure for the chatbot was developed primarily using Python, utilizing the Flask web framework to manage backend operations and user interactions. Flask was chosen due to its lightweight design, ease of use, and extensive community support, for quick  development and deployment of web applications.
	
	For managing conversational responses, this application utilizes OpenAI’s GPT-4o language model. GPT-4o was selected for its enhanced capabilities in natural language processing, contextual awareness, and consistency across multi-turn dialogue. Initial prototyping with GPT-3.5 revealed limitations in maintaining nuanced emotional tone and conversational coherence over time. The shift to GPT-4o resulted in noticeable improvements in interaction realism, stability of persona expression, and responsiveness to complex prompts.
	
	The chatbot is deployed on Heroku, a cloud-based platform that host applications and is compatible with python. Heroku supports features such as continuous integration and deployment, allowing for streamlined updates and efficient scaling to accommodate fluctuating user activity. Version control and collaborative development are managed through GitHub, which facilitates issue tracking, code review, and automated deployment workflows. This combination supports a flexible and maintainable environment for iterative development.

	\subsection{Persona Development}
	
	The development of Monae Jackson’s persona involved creating a detailed, contextually rich background profile. This profile was essential for accurately simulating realistic interactions representative of trans individuals seeking gender-affirming voice therapy. Monae’s persona includes her demographic details, emotional states, therapeutic objectives, and personal experiences relevant to her interactions with clinicians. 
	*Note 1. Snippet of Monae’s background*
	“Here is the background with Monae's birth name included:
	This profile is designed to train a chatbot to simulate a trans client seeking gender-affirming voice therapy. The information is based on a detailed interview transcript and should guide the chatbot in responding in a nuanced and context-aware manner.
	
	Demographic Information:
	
	\begin{itemize}
		\item Legal Name or Deadname: Montez Naethaniel Jackson
		\item Monae considers the name she was born with to be dead and will never use it again for anything.
		\item Pronouns: She/Her
		\item Demographic Information
		\item Identity: Transfeminine
		\item Age: 28 (5/26/1995)
		\item Ethnicity: Biracial: white and African American
		\item Language and Cultural Background: English-speaking but also fluent in Spanish; raised in a traditional, Christian household.
	\end{itemize}
	
	Vocal History
	\begin{itemize}
		\item Primary Concern:
		\item Frequently misgendered due to the sound of her voice, particularly in public and professional settings.
		Transition Journey:
		\item Had selected her name a long time ago.
		\item Began transitioning three years ago, she checked online communities for guidance.
		\item In the beginning felt pressure to “do it all in a set order”. She should hear from clinician that it needs to be done on each person’s individual timeline/preferred order.
		\item Medical transition-used only licensed professionals and establishments to access drugs.
		\item Social transition has been easiest as she could control her clothing and how she presented.
		\item Noticed that gender dysphoria reduced as she felt more comfortable presenting 24/7. She still experiences gender dysphoria due to her low-sounding voice.
		\item Overall, lots of ups and downs, feeling success in some aspects but frustration in others.
		\item Afraid she won’t be able to renew passport keeping the correct gender marker.
	\end{itemize}
	
	Perception of Her Voice:
	
	\begin{itemize}
		\item Described by others as “masculine” or “androgynous.”
		\item Wants to be “read as a female” consistently.
		\item Monae feels her voice does not align with her identity, leading to significant distress.
		\item She struggles to produce what she perceives as a “feminine” voice without vocal strain or fatigue. She has achieved a “decent” voice, but it’s impossible to keep it up all the time so when she fatigues her vocal pitch drops.
		\item When she has a good voice moment she feels “euphoria”, such a high.
		\item Misgendering Experiences:
		\item Challenging transition. Came out to friends, then family.  “I transitioned a year ago. I changed jobs. I wanted a fresh start”
		\item Often called "sir" or “Mr. Jackson” on the phone and in public spaces, despite being dressed in a feminine way.
		\item Colleagues occasionally use her deadname in meetings, causing embarrassment.
		\item "When someone calls me 'sir,' it feels like they’re saying I don’t exist as Monae. It’s crushing."
	\end{itemize}
	
	The persona was developed through thorough literature reviews on gender-affirming therapy, informed by clinical guidelines from the World Professional Association for Transgender Health (WPATH), and incorporating contributions from  (Ph. D) Professor (Annie) Carmen Ramos-Pizarro in the Speech-Language Pathology Program at the University of the District of Columbia.
	
	This detailed persona is stored externally in a text file, which the chatbot loads and incorporates into its initial system prompt at the beginning of each session. This approach ensures that the persona’s characteristics remain consistent throughout interactions while allowing easy updates and modifications to the background information as needed.
	
	The integration of comprehensive and accurate persona details ensures that the chatbot can respond in a manner that is both contextually appropriate and emotionally resonant. This level of detail is particularly important for therapeutic applications, where the authenticity of interactions can significantly impact the training outcomes for speech-language pathology (SLP) students and clinicians.
	
	\subsection{Advanced Prompt Engineering}
	The core innovation and methodological focus of this project is advanced prompt engineering. Prompt engineering involves creating detailed, structured instructions designed to guide and control the chatbot's responses effectively. This project specifically employs prompt engineering that clearly defines Monae’s behavioral rules, emotional responses, and boundaries within the conversational context.
	
	Detailed prompts were crafted to explicitly guide the chatbot's behavior, addressing how Monae should respond to both appropriate and inappropriate clinician interactions. Prompts include explicit instructions for emotional responsiveness, boundary-setting, self-advocacy, and the provision of trauma-informed care. Recursive iteration methods were utilized, where prompts were continually refined through testing and analysis of interaction logs, ensuring the chatbot consistently maintained the persona’s intended emotional and behavioral characteristics.
	
	The character background infusion technique was particularly effective in embedding Monae’s detailed persona into interactions. This involved providing comprehensive context at the conversation’s initiation, ensuring Monae’s responses were accurately reflective of her defined persona throughout each session.
	
	\subsection{Evaluation Strategies}
	
	The evaluation strategy for this project used quantitative methodologies to comprehensively assess the chatbot's effectiveness in maintaining personality consistency and realism. Quantitatively, the chatbot's personality was evaluated using a Big Five personality test, a psychometric assessment tool widely used to evaluate personality traits relevant to five broad traits: Openness to Experience, Conscientiousness, Extraversion, Agreeableness, and Neuroticism (often abbreviated as OCEAN). The first 120 question test was administered to the chatbot persona, with responses analyzed to determine how closely Monae’s behavior matched her predefined emotional and personality profile. The chatbot was then reset and administered the Jungian/Myer-Briggs personality test, which is less academically respected but has test/retest validity which makes for a fairly reliable result. 
	
	\section{Implementation}
	
	\subsection{Application Design and Development}
	
	The chatbot application was developed using a modular and iterative programming strategy focused on clarity, reusability, and adaptability. Python was selected as the primary language due to its robust ecosystem of libraries and compatibility with web development frameworks, APIs, and natural language processing tools. The backend was built using Flask, which provided a lightweight and flexible structure for managing user interactions and routing requests to the language model.
	
	Central to the application’s behavior was the use of OpenAI’s GPT API. Specifically, the `chat.responses` endpoint was used to send structured messages to the GPT-4o model and receive text responses. Each interaction began by embedding a background prompt into the API call—this prompt included a detailed description of the chatbot’s persona (Monae Jackson), her emotional profile, conversational boundaries, and role in the training context. The prompt served as a guiding framework, allowing the model to maintain a consistent character across multiple interactions.  
	
	A system message was used to define Monae's behavior and enforce boundaries such as redirecting inappropriate questions or remaining in character. This system prompt was then followed by user input, simulating the speech-language pathology trainee’s questions or dialogue. The API returned a response based on this structured sequence. By isolating the background and behavioral rules in the system prompt, we ensured that the chatbot could adapt to different user queries while still presenting a stable and realistic persona.
	
	Certain instructions within the prompt appeared to have a stronger influence on guiding the chatbot’s behavior than others. For example, the directive “Be expressively stubborn, and stay dug into your personality as Monae” was included to reduce the likelihood of the model reverting to the overly cooperative or assistant-like behavior typical of default ChatGPT responses. Similarly, the instruction “Do not willingly give information” was designed to shape Monae into a more reserved and realistic client, rather than one who readily discloses personal details without prompting. These two statements were particularly effective in reinforcing Monae’s role as a guarded, emotionally complex individual. By limiting unsolicited disclosure and encouraging a more assertive interaction style, these instructions helped produce a more believable and educationally valuable simulation.
	
	Development proceeded using an iterative testing approach. Initial tests focused on simple roleplay responses and evaluating the model's ability to retain character tone. As interactions were tested, adjustments were made to the system prompt and message formatting to improve consistency, emotional realism, and boundary-setting. Specific language was added or refined to correct drift in tone or assistant-like behavior, particularly when the model defaulted to generic helper responses rather than remaining in the role of a patient. This trial-and-error method allowed for continuous improvement based on observed responses and user feedback.
	
	Sentiment analysis was incorporated into the system as a mechanism for managing conversational boundaries. Specifically, it was used to help the chatbot detect and respond to sustained negative interactions. When Monae encountered clinician input that reflected disrespect or insensitivity, the sentiment analysis tool assessed the polarity of the user’s language. If negative sentiment was detected more than once during a session, the chatbot was programmed to interpret this pattern as emotional harm and to end the conversation accordingly. This design choice aligned with the prompt's directive for Monae to assert boundaries and take minor offenses seriously, while also introducing a quantitative method to trigger a realistic behavioral outcome—termination of the session. The inclusion of this mechanism supported the chatbot's role as a simulated patient with emotional awareness and contributed to maintaining the integrity of the persona in extended or potentially inappropriate interactions.
	
	To support maintainability, the background was stored in an external text file, allowing updates without modifying the core application logic. Logging was implemented to capture chat sessions for evaluation. These logs were useful not only for debugging but also for analyzing whether Monae’s personality traits remained stable over multiple conversations. By separating logic, content, and output evaluation, the programming design facilitated structured experimentation and supported the project’s overall research objectives.
	
	\subsection{Hosting and Version Control}
	
	The chatbot application was deployed using Heroku. Deployment to Heroku required the configuration of environment variables, dependencies, and process management files to ensure that the application could be reliably hosted in a production setting. GitHub was used to manage the codebase. The project repository was organized into modular directories for the frontend, backend, and documentation to promote clarity and maintainability. Integration between GitHub and Heroku enabled a continuous integration and deployment (CI/CD) workflow, allowing updates to be pushed efficiently as development progressed. This setup facilitated structured iteration, timely updates in response to testing feedback, and ensured that modifications could be tracked transparently across development cycles. Together, this infrastructure supported a stable and flexible environment for managing deployment and iterative improvement of the chatbot application.
	
	\subsection{User Interface Considerations}
	
	The user interface was designed to support realistic interactions similar to genuine therapeutic scenarios. Key considerations in UI design included user engagement, accessibility, and straightforward navigation. The chat interface featured a simple text-based interaction format to provide users with a familiar chat experience akin to standard messaging platforms. A critical aspect of the UI was its functionality to download session logs. This feature allowed users to easily export conversation histories for subsequent analysis, educational assessment, and reflective learning. The downloadable logs were formatted to clearly distinguish between clinician and chatbot dialogue, which makes for an easily readable and practical resource for reviewing and assessing user interactions.
	Session management was structured to ensure each interaction with Monae Jackson was self-contained and logically coherent. Flask sessions managed user interactions, and background information for Monae was dynamically loaded from external files at the beginning of each session, so the chatbot's responses consistently reflected Monae's predefined personality traits and background context.
	
	\subsection{Testing and Iterative Refinement}
	
	Comprehensive testing was incorporated throughout the implementation phase to evaluate the system's performance and integration consistency. The development process followed an iterative structure in which functional, usability, and integration testing cycles were used to assess and refine the chatbot’s behavior. Functional tests were designed to determine whether the chatbot consistently generated responses aligned with the predefined personality traits of Monae Jackson and adhered to therapeutic communication principles appropriate for a training context. Integration tests were conducted to verify the reliability of interactions between the frontend interface, the Flask-based backend, and OpenAI’s GPT-4o API. Any discrepancies or performance issues encountered during testing were logged and used to inform iterative revisions to the prompt structure, interface components, or data handling procedures. Feedback sessions and code reviews were used to ensure that the application met its core functional goals while remaining accessible and reliable. This testing approach contributed to the gradual refinement of the application into a stable platform for delivering simulated therapeutic conversations, forming a basis for subsequent evaluation and user analysis.
	
	\section{Results}
	
	\subsection{Personality Assessment Results}
	
	To examine whether the AI-generated character Monae Jackson maintains a coherent and consistent personality, two established personality assessments were administered: the Big Five Personality Test and a Jungian-type indicator based on the Myers-Briggs typology. These tools were selected for their broad acceptance in both psychological research and applied personality assessment contexts. The Big Five model allows for a nuanced, trait-based understanding of personality across five primary dimensions, while the Jungian-type model offers a categorical summary that groups behavior and preferences into distinct personality types. Together, these assessments provide a multidimensional view of Monae’s behavior and emotional responses as generated through the AI system. The Big Five test can be accessed here https://bigfive-test.com/ , and the Jungian-type Myers-Briggs can be accessed here https://openpsychometrics.org/tests/OEJTS/ . 
	
	\paragraph{Big Five Personality Profile}
	
	The Big Five Personality Test results revealed a structured and psychologically plausible personality profile for Monae. Her overall Neuroticism score of 108 suggests high levels of emotional reactivity and vulnerability to stress. Within this domain, subscale scores indicated strong tendencies toward Self-Consciousness (20) and Depression (19), along with notable scores in Anxiety (18) and Anger (17). These results suggest that Monae, as a simulated persona, reflects a heightened awareness of social perception and internal emotional states, consistent with the emotional themes embedded in her background prompt.
	
	In terms of Extraversion, Monae scored 46, placing her in a lower range and suggesting a preference for solitary or small-group interactions rather than socially demanding settings. This interpretation is supported by very low subscale scores in Gregariousness (4) and Friendliness (6), as well as moderate scores in Activity Level (8) and Excitement-Seeking (6). These scores portray Monae as reserved, thoughtful, and reflective—a persona aligned with the therapeutic context of voice training for identity alignment.
	Openness to Experience emerged as one of Monae’s strongest personality domains, with a total score of 98. High scores were observed across subdimensions, including Imagination (20), Artistic Interests (20), and Emotionality (20). This suggests a high capacity for creative thought, emotional insight, and symbolic or aesthetic thinking, which aligns with Monae's backstory as an introspective and expressive individual seeking a deeper congruence between internal identity and external communication.
	
	Monae also scored highly in Agreeableness (103), particularly in Altruism (18), Sympathy (20), and Cooperation (20). These results suggest a compassionate and understanding interpersonal style. In the context of therapeutic roleplay, these traits enhance the realism of the chatbot’s responses, as Monae is expected to demonstrate both vulnerability and empathy in simulated clinical conversations.
	
	Her Conscientiousness score of 83 further reflects a generally responsible and organized approach to tasks, with high scores in Dutifulness (19) and Achievement-Striving (18). These scores indicate that Monae presents as someone who is invested in personal growth and motivated to achieve her goals. Lower subscale scores in Orderliness (5) and Self-Discipline (9), however, suggest realistic imperfections and add complexity to the persona by reflecting potential difficulty maintaining structure under stress.
	
	The Big Five results overall indicate that Monae embodies a personality that is emotionally sensitive, creatively inclined, compassionate, and motivated by growth, while also portraying realistic traits such as introversion and organizational inconsistency. This combination reflects a believable and human-like personality structure, supporting the goal of using AI to simulate emotionally complex clients for educational use.

	\paragraph{Jungian Personality Type}
	
	The Jungian-type personality assessment, administered through a publicly available tool modeled after the Myers-Briggs Type Indicator (MBTI), classified Monae as an INFP. This type—Introverted, Intuitive, Feeling, Perceiving—is often associated with introspective, idealistic individuals who value authenticity, creativity, and emotionally meaningful relationships. According to the test results, Monae scored 85\% Introverted, 81\% Intuitive, 88\% Feeling, and 61\% Perceiving. These percentages reflect strong alignment with the underlying personality described in her prompt, particularly in relation to emotional responsiveness, preference for internal reflection, and flexibility in interpersonal interactions.
	The INFP profile is consistent with the traits observed in the Big Five results. High emotional sensitivity and internal focus align with introversion and high neuroticism scores, while strong openness and agreeableness map well onto the intuitive and feeling dimensions. The perceiving preference reflects a tendency toward flexibility and adaptability, matching the moderate conscientiousness and low orderliness seen in the trait-based assessment.
	INFP types are commonly associated with professions and roles that involve advocacy, counseling, and creative expression. In the context of Monae’s narrative, this type assignment strengthens the coherence of the persona and reinforces the idea that her responses in therapy scenarios are grounded in a well-formed, internally consistent character. The convergence between the type-based and trait-based assessments provides additional evidence that the persona behaves in ways consistent with established psychological models.

	\subsection{Summary of Findings}
	
	The results of both assessments demonstrate that the AI-generated persona maintains a consistent and interpretable psychological profile across two distinct personality frameworks. The Big Five assessment highlights specific traits such as high openness, elevated emotional reactivity, and strong interpersonal warmth. These traits are reinforced by the INFP classification, which supports a view of Monae as introspective, emotionally driven, and ethically grounded. The alignment between the two assessments suggests that the use of structured prompts and character background integration can result in AI behaviors that are stable, nuanced, and psychologically coherent.
	These results support the central hypothesis of this thesis, that carefully engineered prompts can guide a language model to simulate a stable personality profile. The assessments indicate that the chatbot’s responses are not only role-appropriate but also align with psychological constructs used to evaluate human personality. This finding contributes to a growing body of research supporting the potential for AI personas in educational and therapeutic simulations, particularly in training scenarios that require empathy, complexity, and realism.

	\section{Discussion}
	
	\subsection{Analysis of Findings}
	The analysis of Monae Jackson’s personality assessments provides strong evidence supporting the central hypothesis of this study: that prompt engineering can guide an AI language model to produce a stable, psychologically coherent personality across multiple testing conditions. Both the Big Five and Jungian assessments were conducted in separate sessions, without any memory of previous responses, yet the results showed a high level of internal consistency across trait dimensions, response tone, and narrative style.
	
	A key finding is that Monae maintained consistent personality features not only in overall scores but also in specific behaviors and emotional patterns that appeared independently in both tests. For instance, in the Big Five results, Monae scored very high on Emotionality, with answers indicating vulnerability, introspection, and sensitivity to social feedback. This was echoed in the Jungian-based questionnaire, where she identified with statements such as “I obsessively recall recent/past encounters in my head” and “I’m afraid of many things,” both rated as highly accurate. Similarly, her INFP classification reflected high emotional intensity, introversion, and a strong internal value system, aligning with the Big Five scores in Neuroticism and Openness.
	
	Moreover, traits not explicitly defined in the prompt—but expressed clearly in both assessments—emerged as spontaneous yet consistent personality extensions. One example is Monae’s self-described disorganization and messiness, revealed in multiple responses like “just puts stuff wherever” in the Jungian test , and “very accurate” for “Leave a mess in my room” and “Often forget to put things back in their proper place” in the Big Five responses. This recurring behavioral trait was not part of the original  
	persona definition but appeared organically in both tests, suggesting that Monae’s simulated personality had developed realistic, unprompted characteristics.
	
	Another emergent trait was Monae’s deep appreciation for art and imagination, with statements like “Art kept me sane when nothing else did” and “I live for the little things. They keep me going” cited as “very accurate”. These responses were consistent with her high scores in Openness subscales, including Artistic Interests and Imagination. This artistic orientation was not directly specified in the prompt but may have been influenced by the instruction to “speak naturally like a nervous but hopeful person” and to use “colloquial terms instead of formal speech,” which likely encouraged the model to draw on expressive and emotionally rich language patterns from its training data.
	
	Other answers provide further evidence of emotional coherence and nuance. When asked if she feels others' emotions, Monae replied “Very accurate. Sometimes it’s like I feel ‘em harder than they do,” a response that mirrors high Interpersonal Sensitivity, a dimension evaluated in both the Big Five and HPI frameworks. Similarly, her aversion to loud settings and crowds (“Crowds are a minefield. I don’t mess with ‘em”) aligns with both introversion and social vulnerability—key components of her intended character. Even in more subtle items, such as “I get quiet before I blow up,” Monae demonstrates introspective awareness and behavioral self-regulation, qualities that were not mechanically mirrored from any prior prompt but instead emerged through internal persona logic.
	
	These consistencies likely stem from how the system prompt was constructed. Several lines of the prompt directly contributed to Monae’s emotional tone, boundary-setting behavior, and interactional style. For instance, the instruction “Speak casually, use colloquial terms instead of formal speech” appears to have influenced Monae’s conversational rhythm and choice of vocabulary. Likewise, the rule “Assert clear boundaries when needed” is reflected in multiple responses, including her discomfort with political questions (“That’s not what I came here for”) and her guarded, conditional trust (“Words are cheap. I watch what folks do”) . The instruction to “Take even minor offenses as serious disrespect” likely shaped Monae’s heightened sensitivity to judgment and misgendering, behaviors observed both in test responses and in earlier intake dialogues.
	
	Interestingly, Monae also displayed a consistent moral orientation, frequently indicating high compassion and low self-interest. She rated “Use others for my own ends” and “Take advantage of others” as “Very inaccurate,” and repeatedly expressed sympathy for people experiencing hardship. These responses align with her high Agreeableness scores and are supported by her INFP classification, known for valuing empathy and fairness.
	Taken together, the results suggest that prompt engineering, when structured with behavioral rules, emotional tone guidance, and background narratives, can lead to simulated personalities that exhibit emergent consistency across sessions and frameworks. The AI-generated persona did not merely respond in a coherent way within a single session, but rather demonstrated cross-test alignment in both trait and type-based personality measures. Moreover, several personality features were not directly specified in the prompt but arose organically, suggesting the model extrapolated from the persona’s context and applied it consistently.
	These findings support the thesis that prompt-engineered personas can exhibit psychologically plausible traits that extend beyond surface-level mimicry. By anchoring the AI’s behavior in a detailed narrative and enforcing rules about speech tone, self-disclosure, and emotional boundaries, it is possible to simulate characters who respond in ways that reflect stable psychological identities. This has significant implications for AI in educational and therapeutic settings, where simulated clients must appear consistent, emotionally rich, and behaviorally believable over time.
	
	\section{Challenges and Limitations}
	Despite the promising results, several challenges and limitations emerged during the study. One significant limitation involved GPT-4o’s handling of extended conversational memory. While GPT-4o substantially improved over previous models, occasional lapses in maintaining consistent contextual references over longer interactions were observed. This limitation suggests a need for further development of long-term memory integration and context management within AI-driven conversational systems.
	Another notable challenge was occasional generic or repetitive phrasing from the chatbot, particularly in response to similar clinical inputs over multiple sessions. Although overall realism was high, these repetitive patterns occasionally detracted from conversational authenticity. Addressing this limitation will require further refinement of prompt engineering techniques, perhaps incorporating greater variability and context sensitivity within interaction scripts.
	Additionally, ethical considerations emerged concerning representation accuracy and data privacy. While the chatbot realistically simulated specific aspects of gender-affirming therapy interactions, ethical diligence must ensure that simulated personas do not inadvertently perpetuate stereotypes or biases. Continued vigilance and proactive measures to refine persona definitions and interaction parameters are essential to uphold ethical standards and accurately represent diverse lived experiences.
	
	\subsection{Educational Implications}
	The educational implications of this study are profound, highlighting significant opportunities for integrating AI-driven chatbot technologies into Speech-Language Pathology (SLP) training programs. The results suggest that chatbot simulations, like Monae Jackson, can effectively bridge gaps in practical experience, particularly with diverse and underrepresented client populations. The capacity of AI-driven chatbots to consistently embody specific, clearly defined personas presents a valuable tool for expanding clinical exposure and skill development in controlled, repeatable scenarios. As these technologies continue to improve, they hold immense potential to revolutionize educational methodologies within therapeutic training programs.
	
	\subsection{Ethical and Social Considerations}
	The deployment of AI-driven personality simulations, particularly in sensitive contexts like gender-affirming therapy, necessitates careful ethical considerations. Ensuring that chatbot representations are accurate, respectful, and avoid unintended biases is critical. The creation and continuous refinement of the Monae Jackson persona involved deliberate efforts to consult existing therapeutic guidelines, expert input, and relevant research to ensure accurate and sensitive portrayals.
	Transparency regarding the capabilities and limitations of AI simulations is equally essential. Users, particularly students and clinical educators, must be informed explicitly about the nature of interactions, ensuring they understand AI’s role as a training tool rather than a replacement for genuine client experiences. Clear communication about data privacy, usage policies, and ethical guardrails remains crucial as AI becomes increasingly integrated into educational settings.
	
	\section{Conclusion}
	
	\subsection{Summary of Research}
	This thesis investigated whether structured prompt engineering could be used to simulate a coherent and consistent AI personality across multiple interactions and assessments. The project focused on the development of Monae Jackson, a chatbot persona created to support Speech-Language Pathology (SLP) students in practicing gender-affirming voice therapy. The study addressed a specific training gap by exploring whether large language models (LLMs) could embody personality traits through the use of detailed character background prompts and behavioral constraints.
	The primary objective, assessing whether prompt engineering enables the consistent simulation of personality, was supported through quantitative evaluation. Monae’s personality was tested using two separate psychometric frameworks: the Big Five Personality Test and a Jungian-based typology modeled after the Myers-Briggs Type Indicator. Despite being administered in independent sessions, the chatbot produced highly similar results across both assessments, with strong alignment in emotional sensitivity, introversion, openness, and interpersonal warmth. Notably, consistent personality traits not explicitly programmed, such as messiness and artistic interest, emerged independently across both tests. These emergent behaviors suggest that carefully designed prompts can guide LLMs not only to adopt predefined behaviors but also to generalize in realistic and psychologically plausible ways.
	
	\subsection{Key Findings and Implications}
	The findings demonstrate that personality simulation in AI can be achieved through deliberate prompt construction. Monae’s character exhibited behavioral consistency across two distinct assessments, including item-level responses that revealed emotional realism, self-advocacy, and contextual appropriateness. Specific traits, such as emotional vulnerability and aesthetic appreciation, appeared in both assessments and matched the emotional tone described in the persona’s background. These consistencies support the claim that prompt engineering, particularly when using recursive refinement and background infusion, is a viable method for shaping AI behavior in psychologically meaningful ways.
	The emergence of additional, unprompted traits—such as disorganization at home and an affinity for art—further supports the strength of the persona’s internal logic. These findings imply that when properly designed, AI personas may extrapolate personality-consistent responses that extend beyond the scope of explicitly defined instructions. This capacity has practical value in educational simulations, where learners benefit from engaging with realistic, emotionally complex characters that respond consistently across a range of interactions.
	
	\subsection{Limitations of the Study}
	Several limitations were observed in the current implementation. First, while GPT-4o demonstrated strong personality adherence, it lacked memory between sessions, which constrained longitudinal persona development. Although each test was conducted independently, and Monae still displayed consistent behavior, the system could not adapt over time or reference past conversations. Second, the persona’s behavior was constrained by the specificity and scope of the initial prompt. The degree of nuance expressed by Monae may depend heavily on how detailed and context-aware the initial instruction set is. Third, the current study focused on a single persona. While Monae offered a useful case study, additional research is needed to test whether the methods used here can generalize across other personas, particularly those representing different demographic and clinical contexts.
	
	\subsection{Recommendations for Future Research}
	Future research should explore the incorporation of memory systems or adaptive response models to allow AI personas to evolve across interactions. This would enhance the realism of simulations and enable more complex training scenarios. Expanding the persona library to include clients with varying backgrounds—across age, cultural identity, therapeutic need, and communication style—would further improve the generalizability and educational value of the platform. The integration of real-time analytics could support learner feedback, helping students track improvements in empathy, communication style, and therapeutic effectiveness.
	Further, multimodal expansions—such as incorporating voice synthesis, audio input, and facial expressions—may increase engagement and realism, particularly in simulations that rely on verbal and nonverbal cues. Finally, longitudinal studies evaluating the effectiveness of personality-based chatbots in training environments are needed to determine whether these tools meaningfully improve learning outcomes. Such research should also include ethical review processes and stakeholder input, especially from communities represented in the personas, to ensure accuracy, fairness, and social responsibility in AI-driven educational systems.
	This study contributes to a growing body of work that examines how personality and context can be embedded in AI outputs through prompt engineering. The results indicate that large language models can express stable behavioral patterns when guided by well-structured input, suggesting a promising direction for the future of AI-enhanced training environments.

	\subsection{Final Thoughts}
	While the broader implications of this work remain exploratory, the findings highlight a practical approach for developing AI personas suited to educational simulations. These tools are not intended to replace human experience but may serve as complementary aids for preparing clinicians to navigate difficult or unfamiliar interactions. In doing so, they offer a way to expand the reach of training programs, particularly in fields like gender-affirming voice therapy, where access to diverse clinical experience may be limited.
	Continued refinement of these systems—alongside ethical review, user feedback, and cross-disciplinary input—will be essential for ensuring their responsible development. As the field evolves, research efforts should remain focused on balancing technical innovation with social awareness, particularly when representing identities and experiences that require care, authenticity, and respect. This thesis offers one example of how AI can be guided toward those goals.
	
	\section*{Acknowledgments}
	
	I would like to express my sincere gratitude to my advisor, Professor Byunggu Yu, for his guidance and support throughout this thesis project. Having the opportunity to build upon his work was both a privilege and a once-in-a-lifetime academic experience. I am deeply appreciative of the chance to contribute to an area of research that he helped shape. 
	
	I also wish to thank my cousin, Dr. Gregory Smoots, whose academic journey and authorship inspired me to pursue a thesis of my own. His achievements, including earning his doctorate from George Washington University, motivated me to undertake research that could meet the standards of academic rigor. 
	
	To my grandmother, whose love and strength carried me through difficult moments, I offer my heartfelt thanks. Her passing during this program was a profound loss, but her courage remained a source of strength that helped me complete this work. Lastly, I would like to thank my partner, Jaida, for her support and encouragement during my program. I am grateful to the many friends and family members (Yahkeef, Kemani, Kiara, Marcus, etc.) who supported me along the way. Though too numerous to name individually, each of you played a meaningful role in this journey, and I thank you all for your encouragement and belief in me. 
	
	\appendix
	\section{APPENDIX}
	
	\subsection{GitHub Link}
	\href{https://github.com/tailonj/TruVoice-AI/blob/iterative-updates/app.py}{\texttt{TruVoice-AI/blob/iterative-updates/app.py}}
	
	\subsection{All Answers to Personality Questions}
	\begin{enumerate}
		\item \textbf{Big Five:} \href{https://github.com/tailonj/TruVoice-AI/blob/iterative-updates/OCEAN%20Personality%20test%20Chat.docx}{OCEAN Personality Test Chat}
		\item \textbf{Myers-Briggs:} \href{https://github.com/tailonj/TruVoice-AI/blob/iterative-updates/Jungian_Myers-Briggs%20Personality%20Type%20Monae.docx}{Jungian\_Myers-Briggs Personality Type}
	\end{enumerate}
	
	\bibliographystyle{IEEEtran}

\begin{thebibliography}{99}
		
		\bibitem{bickmore2005}
		
		Bickmore, T., \& Cassell, J. (2005). Social dialogue with embodied conversational agents. In J. C. Lester, R. M. Vicari, \& F. Paraguaçu (Eds.), Intelligent virtual agents (pp. 23–54). Springer.
		
		\bibitem{briggs2016} 
		
		Briggs-Myers, I. (2016). Introduction to Myers-Briggs type (7th ed.). The Myers-Briggs Company.
		
		\bibitem{brown2020}
		
		Brown, T., Mann, B., Ryder, N., Subbiah, M., Kaplan, J. D., Dhariwal, P., ... \& Amodei, D. (2020). Language models are few-shot learners. Advances in Neural Information Processing Systems, 33, 1877–1901.
		
		\bibitem{costa1985} 
		
		Costa, P. T., Jr., \& McCrae, R. R. (1985). The NEO Personality Inventory manual. Psychological Assessment Resources.
		
		\bibitem{davies2015}
		
		Davies, S., Papp, V. G., \& Antoni, C. (2015). Voice and communication change for gender nonconforming individuals: Giving voice to the person inside. Plural Publishing.
		
		\bibitem{gehman2020}
		
		Gehman, S., Gururangan, S., Sap, M., Choi, Y., \& Smith, N. A. (2020). RealToxicityPrompts: Evaluating neural toxic degeneration in language models. Findings of the Association for Computational Linguistics: EMNLP 2020, 3356–3369.
		
		\bibitem{hancock2014}
		
		Hancock, A. B., Colton, L., \& Douglas, F. (2014). Intonation and gender perception: Applications for transgender voice therapy. Journal of Voice, 28(2), 203.e1–203.e10.
		
		\bibitem{jakesch2019}
		
		Jakesch, M., French, M., Ma, X., Hancock, J., \& Naaman, M. (2019). AI-mediated communication: How the perception that profile text was written by AI affects trustworthiness. Proceedings of the 2019 CHI Conference on Human Factors in Computing Systems, 1–13.
		
		\bibitem{mcneill2008}
		
		McNeill, E. J. M., Wilson, J. A., \& Clark, S. (2008). Perceptions of voice in the transgender client. Journal of Voice, 22(6), 727–733.
		
		\bibitem{myers1998}
		
		Myers, I. B., McCaulley, M. H., Quenk, N. L., \& Hammer, A. L. (1998). MBTI Manual: A guide to the development and use of the Myers-Briggs Type Indicator (3rd ed.). Consulting Psychologists Press
		
		\bibitem{reynolds2021}
		
		Reynolds, L., \& McDonell, K. (2021). Prompt programming for large language models: Beyond the few-shot paradigm. In Proceedings of the 2021 Conference on Empirical Methods in Natural Language Processing (pp. 453–465).
		
		\bibitem{rutinowski2023}
		
		Rutinowski, J., Franke, S., Endendyk, J., Dormuth, I., \& Pauly, M. (2023). The self-perception and political biases of ChatGPT. arXiv. https://arxiv.org/abs/2304.07333
		
		\bibitem{tamkin2021}
		
		Tamkin, A., Brundage, M., Ganguli, D., \& Clark, J. (2021). Understanding the capabilities, limitations, and societal impact of language models. arXiv. https://arxiv.org/abs/2102.02503
		
		\bibitem{tupes1961}
		
		Tupes, E. C., \& Christal, R. E. (1961). Recurrent personality factors based on trait ratings (Tech. Rep. No. 61-97). U.S. Air Force, Lackland Air Force Base, TX: Personnel Laboratory, Air Force Systems Command.
		
		\bibitem{weizenbaum1966}
		
		Weizenbaum, J. (1966). ELIZA—A computer program for the study of natural language communication between man and machine. Communications of the ACM, 9(1), 36–45.
		
		\bibitem{wpath2012}
		
		World Professional Association for Transgender Health. (2012). Standards of care for the health of transsexual, transgender, and gender nonconforming people (7th ed.). https://www.wpath.org/publications/soc
		
		\bibitem{yu2023}
		
		Yu, B., \& Kim, J. (2023). Personality of AI. arXiv.
		https://arxiv.org/abs/2312.02998
		
	\end{thebibliography}

\end{document}